\documentclass[12pt,fleqn]{iopart}

\expandafter\let\csname equation*\endcsname\relax
\expandafter\let\csname endequation*\endcsname\relax
\usepackage{amsmath}
\usepackage{amssymb}
\usepackage{graphicx}
\usepackage{citesort}

\def\Int#1#2#3{\int_{#1}^{#2}\!\!\!{\rm d}#3\,}
\def\fig#1{Fig.\,\ref{#1}}

\def\vec#1{\mathbf{#1}}

\begin{document}

\title[]{Slow electrons from clusters in strong Xray pulses}

\author{A Camacho Garibay$^{1}$, U Saalmann$^{1}$, and J M Rost$^{1,2}$}
\address{$^{1}$Max-Planck-Institut f\"ur Physik komplexer Systeme\\ 
N\"othnitzer Stra{\ss}e 38, 01187 Dresden, Germany}
\address{$^{2}$PULSE Institute, Stanford University and SLAC National Accelerator Laboratory\\
2575 Sand Hill Road, Menlo Park, California 94025, USA}
\date{\today}

\begin{abstract}
Electrons released from clusters through strong Xray pulses show broad kinetic-energy spectra, extending from the atomic excess energy down to the threshold, where usually a strong peak appears.
These low-energy electrons are normally attributed to evaporation from the nano-plasma formed in the highly-charged clusters.
Here, it is shown that also directly emitted photo electrons generate a pronounced spectral feature close to threshold. Furthermore, we give an analytical approximation for the direct photo-electron spectrum.
\end{abstract}

\pacs{52.20.Fs, 
      36.40.Gk, 
      41.60.Cr, 
      33.80.Wz} 

\maketitle

\section{Introduction}
Recently, there has been increasing interest in slow electrons from photo-driven processes.
While identified in strong-field ionization of atoms \cite{blca+09,quli+09} as well as molecules \cite{dima+14}, the
mechanisms behind the production of slow electrons are very different for atoms in linearly polarized pulses \cite{kasa+12} and molecules in elliptically polarized pulses \cite{dima15}, respectively.
Crucial in both cases is the (single) electron dynamics in the combined potential of the ion left behind and the driving laser field.

Slow electrons can also emerge from soft and even hard Xray pulses. At a first glance this is surprising, since the electronic excess energy $E^{*}$ (which is the photon energy reduced by the binding energy) is typically large, say a few hundred eV up to few keV, depending on the photon energy.
Under such circumstances, the low-energy electrons can occur through non-adiabatic effects in very short pulses, when the pulse length becomes comparable with the orbital period of the bound orbital which is photo-ionized \cite{toto+09,tosa+14}. 

\begin{figure}[b!]
\begin{minipage}[b]{0.37\columnwidth}
\includegraphics[scale=0.75]{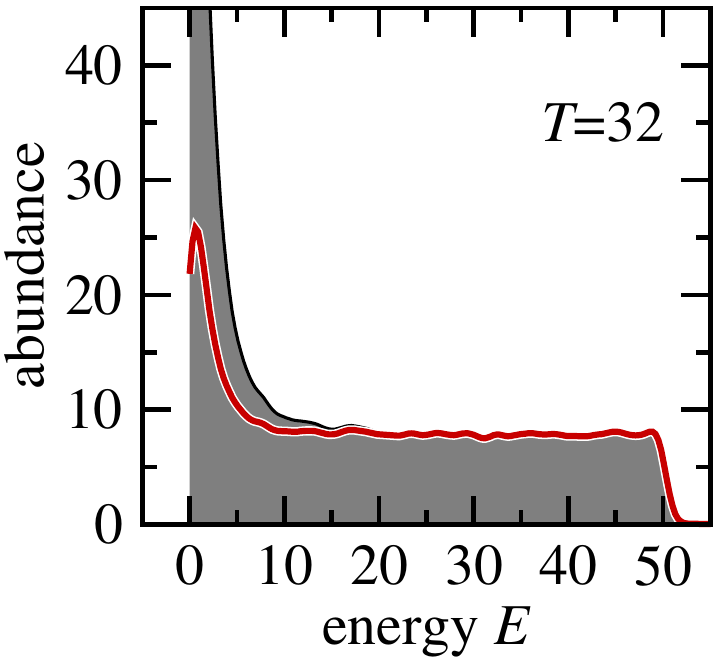}
\end{minipage}\hfill
\begin{minipage}[b]{0.6\columnwidth}
\caption{Electron spectra calculated for a Coulomb complex \cite{gnsa+11} of radius $R=10$ with 10$^{3}$ electrons for an excess energy of $E^{*}\,{=}\,50$ and a pulse duration of $T\,{=}\,32$. Details of the numerical approach are given in section \ref{sec:numerical}.
The full spectrum is shown as darkgray-shaded area, the one for direct electrons only, i.\,e.\ excluding plasma-electrons, as red solid line. 
}
\label{fig:coulcomp1}
\end{minipage}
\end{figure}%
While this effect is again essentially a single-electron phenomenon, 
another very common mechanism to produce slow electrons in intense Xray pulses requires although not collective, yet multiple ionization: Thereby, a complex of ions (either clusters or big molecules) staying behind forms a large background charge
\cite{saro02,mo09,gnsa+09,both+10,arfe10,gnsa+11,arfe11,casa+14}, which reduces the excess energy $E^{*}$.
Hence, these kind of slow electrons can only emerge from large systems, which allow for high charging. In fact, the background charge may be so large that electrons are being trapped even for photons in the keV-range \cite{gnsa+09}. 
The trapping leads to the formation of a so-called nano-plasma,
which thermalizes quickly and consequently evaporates (slow) electrons. 
Typically, the yield of the slow electrons shows an exponential decrease with an energy-scaling constant related to the plasma temperature according to common sense. However, this relation
is tricky for two reasons. Firstly, due to the continuous excitation of electrons into the plasma its state may change considerably during the Xray pulse violating the quasi-stationarity which is necessary
to assign a temperature to the electron spectrum.
Secondly, the photo-ionization process itself gives rise to directly ejected slow electrons.

This is illustrated in \fig{fig:coulcomp1} with the electron spectrum resulting from the illumination of a generic (spherical) cluster with a short pulse
with $T=32$ duration and excess energy of $E^{*}=50$.
The direct electrons (red curve) show a clear peak at low energies revealing that the slow electrons do not only result from the evaporation of the nano-plasma.
We define the direct electrons as those electrons which have a positive energy 
$\vec{p}^{*}{}^{2}/2+W(\vec{r}^{*})>0$ just after the absorption of a photon at time $t^{*}$
at position $\vec{r}^{*}=\vec{r}(t^{*})$ with momentum $\vec{p}^{*}=\vec{p}(t^{*})$. 
The potential energy $W$ involves both the attraction from the ionic background, defined below in section \ref{sec:numerical}, and the electron-electron repulsion.
The initial momentum $\vec{p}^{*}$ is fixed by the excess energy $E^{*}$.

In order to understand the peak in the (numerically obtained) direct-electron yield in \fig{fig:coulcomp1}, we 
will provide in section \ref{sec:analytical} an analytical
derivation of the direct-electron spectrum under the premise that the direct electrons leave the cluster sequentially
and (indirect) plasma electrons remain in the cluster. Thereby, the origin of the slow direct electrons will become clear.
With a surprisingly simple approximation, suggested by the form of the direct-electron spectrum, we can give a fully analytical formula (section \ref{sec:analytical2}).
It is compared in section \ref{sec:numerical} to the numerical direct spectrum, revealing how the indirect plasma electrons influence the direct electrons.

\section{The direct photo-electron spectrum and the origin of slow direct electrons}\label{sec:analytical}
We assume here for simplicity that the system is spherical with a radius $R$ throughout the ionization process.
The light pulse leads to random single-ionization events of atoms within the cluster, where we choose the intensity
such that the system is far from saturation of complete single ionization and the occurrence of any multiple ionization of cluster atoms. More explicitly, if the cluster contains $N$ atoms
and the pulse leads to $Q$ ionization events, then in the end $N{-}Q \approx N$ atoms of the cluster remain neutral.
For the case of sequential ionization the photo-electron spectrum follows from integrating the spectra $P_{q}$ for an instantaneous charge $q$ ranging from $q\,{=}\,0$ (for the initially neutral cluster) to $q\,{=}\,Q$ (the highest possible charge state) 
\begin{equation}\label{eq:gspec}
P(E)=\Int{0}{Q}{q}P_{q}(E).
\end{equation}
The highest charge $Q$ is reached when the cluster potential $V_{q}(r)=[q/2R][3-r^{2}/R^{2}]$
is so deep that absorption of a single photon (with excess energy $E^{*}$) is not sufficient to overcome the threshold$^{\scriptsize\footnotemark}$\footnotetext{Note that higher charge states can be reached when electrons are excited below threshold and the nano-plasma, formed in the process, evaporates \cite{saro02,both+10,yana+13}.}. 
This occurs if $V_{Q}(R)=E^{*}$ which implies $Q=E^{*} R$. 

\begin{figure}[t!]
\centering
\includegraphics[width=0.7\columnwidth]{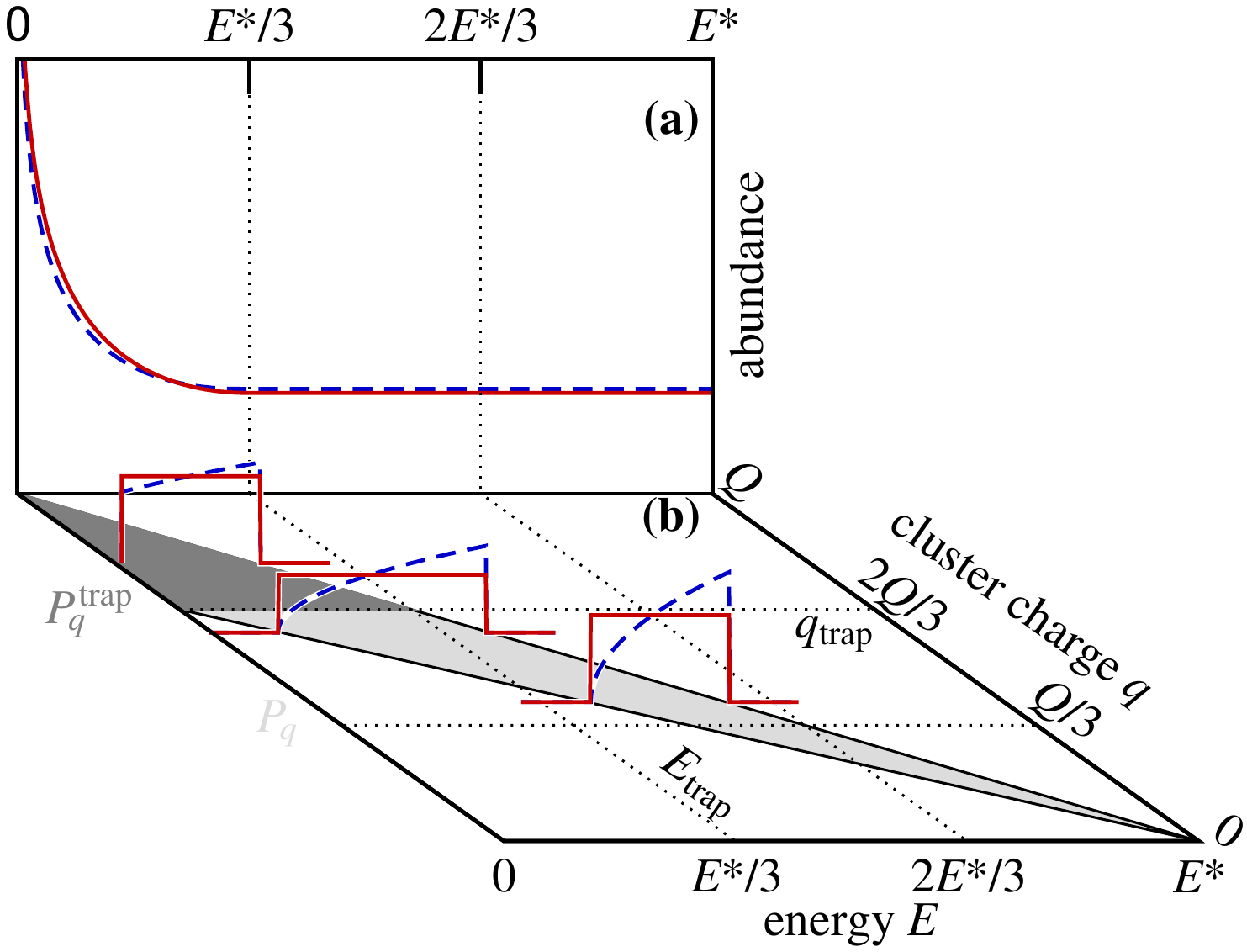}
\caption{Sketch of formation of the photo-electron spectrum
{\bf(a)} Final spectrum as obtained by numerical integration (blue dashed line) of Eq.\,\eqref{eq:gspec} with Eqs.\,\eqref{eq:qspec2} and \eqref{eq:qspec3} and from the analytical approximation (red solid line) according to Eq.\,\eqref{eq:rect};
{\bf(b)} The contribution from a particular charge:
The shaded area shows which charges $q$ contribute to which energy $E$ either according to $P_{q}$ (light-gray) or $P_{q}^{\rm trap}$ (dark-gray), respectively.
Additionally, there are three explicit examples with $q=2Q/5,\,3Q/5,\,4Q/5$ for these distributions according to Eqs.\,(\ref{eq:qspec2},\,\ref{eq:qspec3}) and \eqref{eq:appspec}, respectively.
}
\label{fig:sketch}
\end{figure}

If the cluster potential $V_{q}(r)$ is still shallow enough for all electrons in the cluster to escape by absorbing just one photon, the electron spectrum of a $q$-fold charged spherical cluster is given by \cite{gnsa+13}
\begin{subequations}\label{eq:qspec1}
\begin{equation}
P_{q}(E)=\frac{3}{R^{3}}\Int{0}{R}{r}r^{2}\delta\big(E-E_{q}(r)\big)
\end{equation}
with 
\begin{equation}\label{eq:qspec1b}
E_{q}(r)=E^{*}-\frac{q}{2R}\Big[3-\frac{r^{2}}{R^{2}}\Big]
\end{equation}
\end{subequations}
the final energy of an electron released at a distance $r$ from the centre through the absorption of a photon.
We obtain from Eq.\,\eqref{eq:qspec1} 
\begin{subequations}\label{eq:qspec2}
\begin{align}
P_{q}(E) = {} & \frac{3}{q/R}\sqrt{3-2\frac{E^{*}{-}E}{q/R}}
\qquad\mbox{for}\quad E_{\rm min}(q)\le E\le E_{\rm max}(q)
\\
&\mbox{with}\quad
E_{\rm min}(q)\equiv E^{*}-3q/2R \quad\mbox{and}\quad
E_{\rm max}(q)\equiv E^{*}-q/R 
\end{align}
\end{subequations}
and $P_{q}(E)=0$ elsewhere. Here, $E_{\rm min}(q) = E_{q}(0)$ is the energy from an electron released at the centre ($r=0)$, while an electron from the surface will appear at $E_{\rm max}(q) = E_{q}(R)$, see
Eq.\,\eqref{eq:qspec1b}. 
The two lower blue dashed lines in \fig{fig:sketch}b show as examples $P_{q}(E)$ for $q{=}2Q/5$ and $q{=}3Q/5$, respectively.

Expression \eqref{eq:qspec2} has to be modified when the cluster potential becomes so deep that electrons\,---\,firstly those released at the centre\,---\,are trapped after single-photon absorption. 
This occurs at $q_{\rm trap}=E^{*} 2R/3=2Q/3$.
Since these electrons do not increase the cluster charge, further charging is due to electrons located
closer and closer to the surface.
Hence, for charges $q_{\rm trap}\le q\le Q$ the instantaneous spectra become
\begin{equation}\label{eq:qspec3}
P_{q}(E)\to P^{\rm trap}_{q}(E)=\frac{1}{1-\big[3{-}2\frac{E^{*}}{q/R}\big]^{3/2}}P_{q}(E)
\qquad\mbox{for}\quad 0\le E\le E_{\rm max}(q)
\end{equation}
and $P^{\rm trap}_{q}(E)=0$ elsewhere.
The prefactor in \eqref{eq:qspec3} normalizes the distribution $P^{\rm trap}_{q}$ for any $q$ just as $P_{q}$ above is normalised, i.\,e.,
\begin{equation}\label{eq:norm}
\Int{0}{E^{*}}{E}P_{q}(E)=\Int{0}{E^{*}}{E}P^{\rm trap}_{q}(E)=1.
\end{equation}
One example for $P^{\rm trap}_{q}(E)$ is shown in \fig{fig:sketch}b with the upper blue dashed line corresponding to $q{=}4Q/5$. One can also see from the shaded area in \fig{fig:sketch}b that the restriction of electron energies to the interval $E_{\rm min}(q)\le E(q) \le E_{\rm max}(q)$ implies for the integral \eqref{eq:gspec} a restriction to charges in the interval $q_{\rm min}(E)\le q(E)\le q_{\rm max}(E)$
with
\begin{equation}
q_{\rm min}(E)\equiv\frac{2}{3}[E^{*}{-}E]R
\quad\mbox{and}\quad
q_{\rm max}(E)\equiv[E^{*}{-}E]R,
\end{equation}
which follows directly from Eq.\,(\ref{eq:qspec2}b).
The abundance for a particular energy $E$ finally reads
\begin{subequations}\label{eq:exaspec}
\begin{align}
P(E) & = \Int{q_{\rm min}}{q_{\rm max}}{q}P_{q}(E) && \mbox{for}\quad E^{*}/3\le E\le E^{*}
\\\label{eq:exaspecb}
P(E) & = \Int{q_{\rm min}}{q_{\rm trap}}{q}P_{q}(E)+\Int{q_{\rm trap}}{q_{\rm max}}{q}P^{\rm trap}_{q}(E) 
&& \mbox{for}\quad 0\le E\le E^{*}/3.
\end{align}
\end{subequations}
Equation (\ref{eq:exaspec}a) can be solved analytically and gives the energy-independent value 
$P(E)=3R\big[\sqrt{3}\,\ln\big(2{+}\sqrt{3}\big){-}2\big]$
corresponding to a plateau \cite{gnsa+11}.
Equation (\ref{eq:exaspec}b), on the other hand, does not allow for a compact analytically solution. Therefore, we provide with the blue dashed line in \fig{fig:sketch}a the numerically integrated spectrum. 
One clearly sees an accumulation towards lower energies with a divergence at $E=0$. With $q_\mathrm{min}$ and $q_\mathrm{trap}$
finite, this is due to the second term in \eqref{eq:exaspecb} and may be interpreted as follows: Electrons with energies $E\ge E^{*}/3$ can escape from anywhere in the cluster for any (accessible) charge state, as described by
Eq.\,(\ref{eq:exaspec}a). For electrons with energies $E<E^{*}/3$ this is limited to clusters charged less than
$q_\mathrm{trap}=2Q/3$, cf.\ 1st integral in (\ref{eq:exaspec}b). For clusters charged higher than $q_\mathrm{trap}$ direct electrons come from the outer regions of the cluster with ever decreasing energy as the cluster charge grows beyond $q_\mathrm{trap}$. This part of the spectrum is described by the 2nd integral in (\ref{eq:exaspec}b).

\section{Analytical approximation for the direct photo-electron spectrum}\label{sec:analytical2}

Interestingly, the exact shape of $P_{q}(E)$ is not important for the final spectrum. 
One may choose any form for $P_{q}$. As long as the shape for various values $q$ can be obtained by a simple scaling the final spectrum is a plateau \cite{gnsa+13}.
In order to obtain an analytical expression for all energies we approximate $P_{q}(E)$ with the simplest form possible, namely a constant spectrum between $E_{\rm min}$ and $E_{\rm max}$. The $q$-dependence is then introduced via the normalization
\eqref{eq:norm}. With $E_{\rm min}(q)$ and $E_{\rm max}(q)$ given in (\ref{eq:qspec2}b) this leads to
\begin{subequations}\label{eq:appspec}
\begin{align}
P_{q}(E) & = \frac{1}{E_{\rm max}-E_{\rm min}} =\frac{2R}{q} && \mbox{for}\quad 0\le q\le q_{\rm trap}
\\
P_{q}(E) & = \frac{1}{E_{\rm max}} = \frac{1}{E^{*}-q/R}
&& \mbox{for}\quad q_{\rm trap}\le q\le Q.
\end{align}
\end{subequations}
This distributions are shown in \fig{fig:sketch}b as red solid lines.
They allow for an integration of Eq.\,\eqref{eq:exaspec}
\begin{subequations}\label{eq:rect}
\begin{align}
P(E)&
=2R\,\ln(3/2) 
&& \mbox{for}\quad E^{*}/3\le E\le E^{*}\label{eq:rect1}
\\
P(E)&
= 2R\,\ln\Big(\frac{E^{*}}{E^{*}{-}E}\Big)+ R\,\ln\Big(\frac{E^{*}}{3E}\Big)
&& \mbox{for}\quad 0< E\le E^{*}/3\label{eq:rect2}
\end{align}
\end{subequations}
which is shown as red solid line in \fig{fig:sketch}a. 
It reproduces the spectrum obtained with the blade-shaped instantaneous spectra (blue dashed line in \fig{fig:sketch}a) extremely well. This applies to both, the absolute values of the plateau at large $E$, and the divergent behaviour around $E{=}0$. Note, that the latter is indeed due to the second term in \eqref{eq:rect2} which represents the analytical approximation of the
integral over $P_{q}^\mathrm{trap}$ in \eqref{eq:exaspecb}.

\section{Comparison to numerical results from Coulomb complexes}\label{sec:numerical}
The analytical expressions have been derived under the assumptions that photo-ionization occurs sequentially and that electrons excited to states below threshold remain trapped.
In the following we will assess if and when these assumptions are justified
 by comparing the results from ({\ref{eq:rect}) to those of molecular dynamics calculations without those assumptions in the framework of so-called photo-activated Coulomb complexes \cite{gnsa+11}.
This is a simple model, where electrons are treated as classical particles and ions form a spherical jellium, describing the attractive potential of the charged cluster.
This potential as well as the electron-electron interaction is essential for understanding the formation of the broad electron spectra.
Once activated (i.\,e.\ released with a given energy determined by the excess energy $E^{*}$) the electrons are propagated according to Newton's equations with forces resulting from the jellium potential and the electron-electron interaction.
Thus, in contrast to the description of the previous section, here correlations (collisions) of the electrons are fully taken into account. 
The system is propagated sufficiently long (up to times $t\,{=}\,10^{4}$ for the results presented) before spectra are calculated. 
These are obtained by folding the final (kinetic) energies $E_{j}$ of the electrons with a Gaussian 
\begin{equation}
P(E)=\sum_{j}\exp\big({-}[E_{j}{-}E]^{2}/\delta E^{2}\big)
\end{equation}
of width $\delta E{=}1$.

\begin{figure}[t!]
\centering
\includegraphics[scale=0.75]{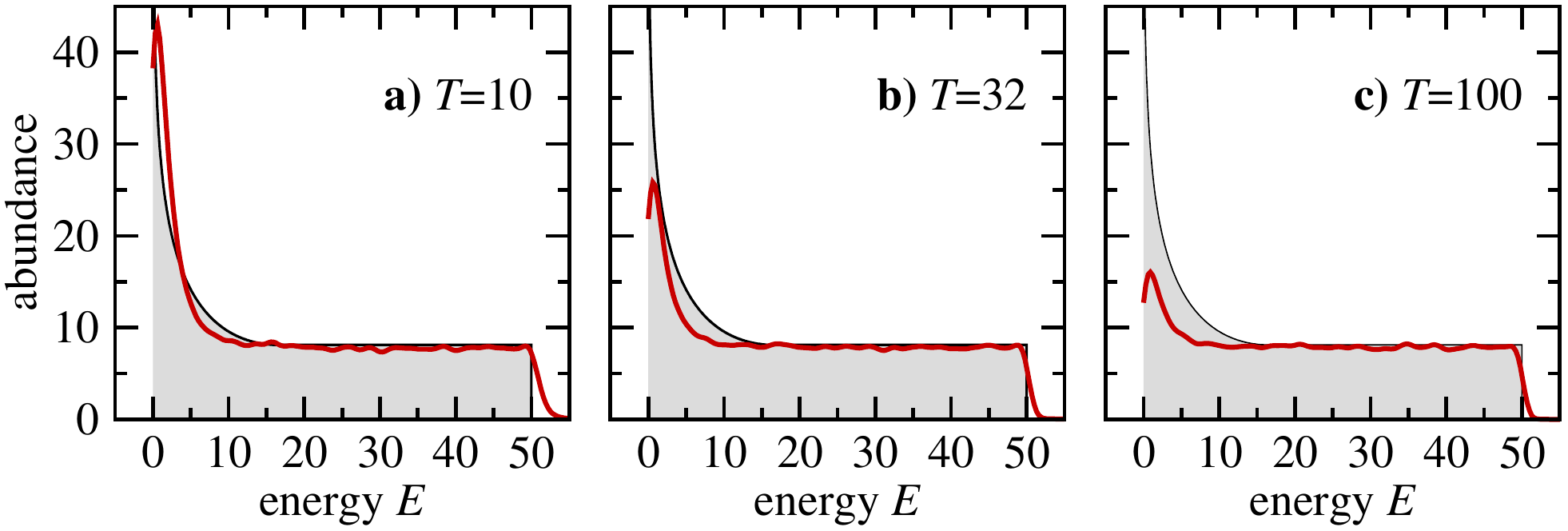}
\caption{Electron spectra (red solid line) for the direct photo-electrons only, i.\,e.\ excluding plasma-electrons, as obtained from Coulomb complexes with 10$^{3}$ electrons for an excess energy $E^{*}\,{=}\,50$ and various pulse durations $T$. 
They should be compared to the analytical expressions \eqref{eq:rect}, which is shown by the gray-shaded areas.}
\label{fig:coulcomp2}
\end{figure}%
Figures \ref{fig:coulcomp1} and \ref{fig:coulcomp2} show such spectra (obtained by averaging over 100 realizations) for a Coulomb complex of radius $R\,{=}\,10$ with 10$^{3}$ electrons and $E^{*}\,{=}\,50$. The photo activation rate is proportional to $\exp\big({-}t^{2}/T^{2}\big)$.
One clearly sees a broad spectrum with a large peak at $E\approx0$, a plateau at $E\,{<}\,E^{*}$ and a cutoff at $E\,{=}\,E^{*}$. These features have been observed \cite{both+08} and discussed \cite{both+10,gnsa+11,arfe10,arfe11} before, interpreting the high-energy part (plateau) as a consequence of the direct photo electrons and the low-energy part with its peak towards threshold as a consequence of the evaporation from the transient nano-plasma. 

However, as already mentioned, \fig{fig:coulcomp1} reveals that also direct electrons, defined as those electrons which have initially enough energy to escape from the 
cluster potential, contribute to the slow-electron peak. Their contribution to the low energy spectrum is even larger in the analytical estimate considering only sequentially emitted electrons (see \fig{fig:coulcomp2}, gray-filled area)
than from the numerically obtained direct electrons (red curve). The reason is that we do not take into account that
initially trapped plasma electrons do eventually leave at a certain rate, dictated by the plasma temperature.
If this rate is faster than the photo ionisation rate, direct photo electrons see an increased background charge reducing 
their yield at low energies since they get trapped. This effect should be least important for very short pulses when the direct electrons leave before plasma evaporation becomes important. However, for very short pulses, the second assumption made for the analytical direct electron spectrum is violated, namely, the sequential ionization: The photo-ionization rate is so large that the direct electrons interact and exchange energy before leaving the cluster. This indicates the onset of massively parallel ionization \cite{gnsa+12}, which is accompanied by high-energy tails at $E\,{\gtrsim}\,E^{*}$ in the spectrum.
Indeed, the red curves from the numerical calculation in \fig{fig:coulcomp2} show these tails in contrast to the sharp cutoff of the analytical spectrum at $E=E^{*}$. 

\section{Summary}
Comparing fully numerical spectra to those from photo electrons only, we have shown that the low-energy peak observed in the photo-electron spectrum of multiple ionization of clusters in strong Xray pulses is not only generated by initially trapped plasma electrons but also by photo electrons directly escaping. An understanding of the origin of slow direct electrons has been made possible by the formulation of the spectrum for the direct electrons alone down to threshold, including a fully analytical approximation\,---\,always under the assumption that the electrons leave the cluster sequentially.
In the future it would be interesting to disentangle direct photo-electron dynamics from plasma-electron dynamics experimentally. This could be done by by exploiting the fact that angular distributions are different for photo-electrons (depending on the shape of the orbitals being ionized) and plasma-electrons (expected to be isotropic) or by using streaking techniques \cite{kr14}.

\vfill\noindent\hrulefill

\end{document}